\documentclass[preprint2]{aastex}

\usepackage{amsfonts}
\usepackage{amsmath}
\usepackage{amssymb}

\begin{document}

\title{GRAVITATIONAL STABILITY OF A CYLINDRICAL PLASMA WITH AN AZIMUTHAL AND AN AXIAL MAGNETIC FIELD}

\author{J. A. McLeman\altaffilmark{1}, C. H.-T. Wang\altaffilmark{1,2} and R. Bingham\altaffilmark{2,3}}
\affil{$^{1}$SUPA Department of Physics, University of Aberdeen, King's
College, Aberdeen, AB24 3UE, UK \\$^{2}$STFC Rutherford Appleton
Laboratory, Chilton, Didcot, Oxfordshire, OX11 0QX, UK \\ $^{3}$SUPA Department of Physics, University of Strathclyde, Glasgow, G4 0NG, UK}

\begin{abstract}
We consider the gravitational stability of a current carrying filamentary cloud in the presence of both axial and azimuthal magnetic fields using a simple analytic model. The azimuthal magnetic field is shown to give rise to a new contribution, dictated by Ampere's law, in the corresponding virial equation for magnetohydrodynamic equilibrium. From this we obtain a computationally inexpensive guidance on the gravitational stability of current carrying filamentary clouds. The approach not only provides a fresh insight into the essential physical mechanisms involved, but also demonstrates clearly that, for sufficiently large and yet astronomically realistic currents, the azimuthal magnetic field can cause filamentary clouds to undergo instability.
\end{abstract}


\section{INTRODUCTION}

One of the fundamental conditions for the onset of the dynamical instability causing the disperse, cold gas to collapse was given by the Jeans condition (Jeans 1902). This gave us a requirement for the instability to occur as an expression of the velocity of a propagating wave through the gas. When the velocity of the wave became imaginary we would no longer see oscillations but an exponential collapse at the boundary instead. This could also be expressed as a minimum, critical wavelength beyond which we would see an instability occurring. When considering these molecular clouds they also arise in cylindrical filamentary mass distributions, such as the \textit{``Nessie''} Nebula (Jackson et al. 2010). Further, infrared dark clouds (IRDCs) with a filamentary structure have been identified by new telescopes such as \textit{Herschel} (Arzoumanian et al. 2011; Hill et al. 2011).

To account for this matter distribution the Jeans stability condition was modified. This modification showed that cylindrical clouds, subject to longitudinal perturbations, can (under specific conditions) break apart into smaller pieces (Dibai 1957), the results of which could be used to explain the creation of star chains. Both of these methods to understand this instability neglect the magnetic field and its influence on the stability of the system. It was not until 1953 when Chandrasekhar and Fermi approached this problem for the first time (Chandrasekhar $\&$ Fermi 1953). They realized that for stationary configurations it is sufficient to consider a less stringent stability condition called the virial equation. The method of analyzing the stability through the virial equation has the advantage of being much simpler than directly tackling often cumbersome partial differential equations, this is often the case when considering a method similar to the derivation of the Jeans condition (Jeans 1928). By considering a plasma, they first derived the virial equation in the most general case including a magnetic energy term. This was done by first considering the Navier-Stokes equation in three dimensions with the addition of magnetic force density terms. By integrating the equation over the volume they obtain an expression involving the energy terms from the different components, i.e, the magnetic energy and the molecular energy. This can be used to represent a condition required for the system to be in equilibrium. Using a similarly fashioned derivation as done in the most general case, they then did the same for cylindrical distributions of matter with a prevailing axial magnetic field. From this a stability condition was reached. It was shown that the average magnetic intensity is required to be less than a critical value, which is dependent on the average mass density of the cylinder, in order to maintain stability. However, Dibai (1958) showed that for longitudinal waves propagating through a cylindrical medium of plasma, an axial magnetic field does not affect the stability of the system.

This previous research focuses on analytical methods by first assuming that the system starts close to gravitational stability and then showing it can fragment. However, the interstellar medium (ISM) is highly dynamical and yet filaments can be shown to form in these environments through the use of numerical simulations (Ntormousi et al. 2011). The gravitational collapse in turbulent molecular clouds is influenced by the magnetic field. It can be shown that magnetic fields cannot prevent a local collapse unless they act as a magnetostatic support (Heitsch et al. 2001).  Despite the turbulence of the ISM it is possible to form pillar like molecular clouds and through numerical simulations it is shown that it can be due to impact ionization from nearby massive stars. Gravitational instability can be induced at the tip of these pillars leading to the formation of low mass stars (Gritschneder et al. 2009). Molecular clouds in general can form in many different shapes and sizes but the background magnetic field can be shown to strongly influence its development. Further numerical simulations show that the direction and strength of this magnetic field is vital in determining the form of the molecular cloud (Heitsch et al. 2008). The properties of clumps of interstellar material formed in the presence of weak magnetic fields with a colliding warm neutral medium is investigated through the use of numerical magnetohydrodynamic (MHD) simulations (Banerjee et al. 2009). It was shown that the magnetic field has an alignment with the velocity but there exist intense fluctuations both in magnitude and direction within the clumps meaning that the magnetic field has become distorted by the turbulence of the gas. Despite this a dynamical collapse of the cloud is still possible. The evolution of giant molecular clouds when subject to a collision with a warm neutral medium streams in the presence of magnetic fields and ambipolar diffusion has been investigated through the use of numerical MHD simulations (V\'{a}zquez et al. 2011). It was shown that only for supercritical or marginal critical streams could reasonable rates of star formation occur. It was also shown that the mass to magnetic flux ratio is a fluctuating function of position within the cloud which they use to imply that a large portion of the cloud's mass may remain magnetically supported against collapse whereas star formation occurs in the non-supported regions.

There have been observations to suggest that some molecular clouds may exhibit helical magnetic fields (Bally 1987), consisting both axial (poloidal) and azimuthal (toroidal) components. It has also been found that a current carrying jet creating an azimuthal magnetic field is a useful model to explain the temporary capture of carbon monoxide (CO) in the spiral galaxy NGC4258 (Krause et al 2007). The presence of the azimuthal magnetic field will influence the stability of the system, which has been addressed by a number of recent papers (Fiege $\&$ Pudritz 2000a, 2000b). The effect of the azimuthal magnetic field on rotating, isothermal clouds has been investigated in (Tomisaka 1991). The main results of these works are obtained through MHD numerical simulations. The gravitational collapse and fragmentation of magnetized filamentary clouds has been investigated and it can be shown that an azimuthal magnetic field can strengthen the cylindrical cloud against fragmentation during collapse (Tilley $\&$ Pudritz 2003). Numerical MHD simulations including turbulence have shown that the magnetic field plays a significant role in the collapse of molecular clouds and the formation of star-forming cores (Tilley $\&$ Pudritz 2007). The influence of the magnetic field on the star formation rate in molecular clouds is profound. Even in the presence of weak magnetic fields the formation of dense cores in the star formation process can be slowed significantly resulting in a longer evolution time (Nakamura $\&$ Li 2011).

The purpose of the present paper is to provide an analytical approach to the gravitational stability of a cylindrical filament of plasma which is subject to both axial and azimuthal magnetic fields. We focus on the essential physical mechanism and aim to provide a computationally inexpensive guidance on the corresponding necessary condition of gravitational stability. Starting from the most general fluid flow equation we derive a new form of virial equation. Specifically, it yields information on how gravitational stability is affected by the presence of azimuthal magnetic field.

\section{THE CONFIGURATION OF A CYLINDRICAL PLASMA WITH MAGNETIC FIELDS}

We shall idealize a dark filamentary plasma as an infinitely long cylinder with radius $R$ of fluid having zero resistivity, along whose axis we allow for a net movement of charge. This sets up an azimuthal magnetic field, just like that of a current carrying wire, in addition to a possible prevailing magnetic field in the axial direction. It is natural to adopt the cylindrical polar coordinates $(r, \phi, z)$ where $z$ denotes the axis of the filament. The cylindrical symmetry requires the all physical quantities involved including the density $\rho$, pressure $p$, velocity $\textbf{v}$ and magnetic field $\textbf{B}$ to be functions of radial distance from the axis $r$ and time $t$. In particular both azimuthal and axial components of the magnetic field, $B_\phi$ and $B_z$ respectively, depend only on $r$ and $t$.

The magnetic fields across the boundary of the cylinder are assumed to match smoothly (continuous and differentiable) so that there is no surface current sheet. Exterior to the cylinder the axial magnetic field $B_z$ is assumed to be zero for simplicity while the azimuthal magnetic field $B_\phi$ may remain non-vanishing.

\section{VIRIAL EQUATION WITH AXIAL AND AZIMUTHAL MAGNETIC FIELDS}

We begin by considering the Navier-Stokes equation for an inviscid fluid in the MHD approximation as follows
\begin{equation}\label{base}
\rho\frac{d \textbf{v}}{d t}
   = -\nabla (p+\frac{|\textbf{B}|^{2}}{2\mu_{0}}) + \rho \nabla V +\frac{(\textbf{B}.\nabla)\textbf{B}}{\mu_{0}}
\end{equation}
where $\textbf{v}$ is the velocity, $t$ is the time, $p$ is the pressure, $V$ is the gravitational potential, and is the $\rho$ density . Here $\frac{d \textbf{v}}{d t}$ denotes the Lagrangian time derivative of $\textbf{v}$. The cylindrical symmetry to be used throughout yields
\begin{equation}\label{lag}
\frac{d \textbf{v}}{d t}=\hat{\textbf{z}} \frac{d v_{z}}{d t} + \hat{\textbf{r}} \frac{d v_{r}}{d t}
\end{equation}
and
\begin{equation}\label{bred}
-\nabla (p+\frac{|\textbf{B}|^{2}}{2\mu_{0}})=-(\hat{\textbf{r}} \frac{\partial}{\partial r} ) (p+\frac{|\textbf{B}|^{2}}{2\mu_{0}})
\end{equation}
in cylindrical polar coordinates.
The second term on the right-hand side of Equation \eqref{base} is a gravitational term, satisfying
\begin{equation}
\oint \textbf{g}.d\textbf{A} = -4\pi G M_{\text{enc}},
\end{equation}
where $\textbf{g}$ is the gravitational acceleration vector with radial only components at radius $r$ , $d\textbf{A}$ is the surface area unit, and $M_{\text{enc}}$ is the mass enclosed by an imaginary cylindrical surface of radius $r$ and length $L$. This yields
\begin{equation}\label{g2}
\textbf{g} = - \frac{2 G \lambda_{\text{enc}}}{r} \hat{\textbf{r}}.
\end{equation}
where $\lambda_{\text{enc}}=M_{\text{enc}}/L$ is the mass per unit length enclosed within radius $r$. For $r > R$ we shall denote the constant  mass per unit length by $\lambda=\lambda_{\text{enc}}$. Thus we see that
\begin{equation}\label{g3}
\rho \nabla V = \hat{\textbf{r}} \rho(\frac{\partial}{\partial r} ) V = - \frac{2 G \lambda_{\text{enc}}}{r} \rho \hat{\textbf{r}}.
\end{equation}

The third term on the right-hand side of Equation \eqref{base} is the magnetic tension term. To evaluate this term, we consider the following general expansion in cylindrical coordinates:
\begin{align*}
&
(\textbf{P}.\nabla)\textbf{Q}=\\
&\;\;\;
(P_r\frac{\partial Q_{r}}{\partial r} + \frac{P_\phi}{r} \frac{\partial Q_r}{\partial \phi}+P_z \frac{\partial Q_r}{\partial z} -\frac{P_\phi Q_\phi}{r})\hat{\textbf{r}} \notag \\
&\;\;\;
+ (P_r \frac{\partial Q_\phi}{\partial r} + \frac{P_\phi}{r} \frac{\partial Q_\phi}{\partial \phi} + P_z \frac{\partial Q_\phi}{\partial z} +\frac{P_\phi Q_r}{r})\hat{\boldsymbol{\phi}} \notag \\
&\;\;\;
+ (P_r \frac{\partial Q_z}{\partial r} + \frac{P_\phi}{r} \frac{\partial Q_z}{\partial \phi} + P_z \frac{\partial Q_z}{\partial z})\hat{\textbf{z}}
\end{align*}
for two arbitrary vector fields $\textbf{P}$ and $\textbf{Q}$. On restricting
both $\textbf{P}$ and $\textbf{Q}$ to be $\textbf{B} = B_\phi(r,t)\hat{\boldsymbol{\phi}}+B_z(r,t)\hat{\textbf{z}}$, the above expansion reduces simply to
\begin{equation}\label{bred2}
(\textbf{B}.\nabla)\textbf{B} = -\frac{B_\phi^{2}}{r}\hat{\textbf{r}}
\end{equation}
See, e.g., Arfken (1985).

Therefore, by substituting relations \eqref{lag}, \eqref{bred}, \eqref{g3} and \eqref{bred2} into Equation \eqref{base}, we see that
\begin{align}\label{flow}
\rho(\hat{\textbf{r}}\frac{d v_{r}}{d t} + \hat{\textbf{z}}\frac{d v_{z}}{d t}) = -(\hat{\textbf{r}} \frac{\partial}{\partial r} ) (p+\frac{|\textbf{B}|^{2}}{2\mu_{0}})\notag \\ - \frac{2 G \lambda_{\text{enc}}}{r}\rho \hat{\textbf{r}}-\frac{B_\phi^{2}}{\mu_{0}r}\hat{\textbf{r}}.
\end{align}
Integrating Equation \eqref{flow} over the cross-section area of the cylinder, we obtain two equations from the resulting $\hat{\textbf{z}}$ and $\hat{\textbf{r}}$ components respectively as follows:

\begin{equation}
 \int_{\phi=0}^{\phi=2\pi} \! \int_{r=0}^{r=R} \rho r\frac{d v_{z}}{d t}\,rd\phi\,dr = 0.
\end{equation}

\begin{align}\label{int1}
\int_{\phi=0}^{\phi=2\pi}  \int_{r=0}^{r=R} \rho r\frac{d v_{r}}{d t}rd\phi dr = \notag \\ -\int_{\phi=0}^{\phi=2\pi}  \int_{r=0}^{r=R} r(\frac{\partial}{\partial r}) (p+\frac{|\textbf{B}|^{2}}{2\mu_{0}})rd\phi dr \notag\\- 2\int_{\phi=0}^{\phi=2\pi}  \int_{r=0}^{r=R} G\lambda_{\text{enc}}\rho rd\phi dr \notag \\ -\int_{\phi=0}^{\phi=2\pi}   \int_{r=0}^{r=R} \frac{B_\phi^{2}}{\mu_{0}}rd\phi dr.
\end{align}
To proceed we shall denote by
\begin{equation}
\int_{\phi=0}^{\phi=2\pi} \! \int_{r=0}^{r=R} \rho\,rd\phi\,dr = \int_{0}^{\lambda}\!\,d\lambda
\end{equation}
where $d\lambda$ is a mass per unit length element. Evaluating the gravitational integral first, we get

\begin{align}\label{use1}
- 2\int_{\phi=0}^{\phi=2\pi} \! \int_{r=0}^{r=R} G\lambda_{\text{enc}}\rho \,rd\phi\,dr=\notag \\ -2\int_{0}^{\lambda}\! G\lambda_{\text{enc}}\,d\lambda = -G\lambda^{2}.
\end{align}
We evaluate the first integral on the right-hand side of Equation \eqref{int1} by considering the angular integral,

\begin{align}
-\int_{\phi=0}^{\phi=2\pi} \int_{r=0}^{r=R} r(\frac{\partial}{\partial r}) (p+\frac{|\textbf{B}|^{2}}{2\mu_{0}})rd\phi dr = \notag \\ -2\pi \int_{r=0}^{r=R} r^{2}(\frac{\partial}{\partial r}) (p+\frac{|\textbf{B}|^{2}}{2\mu_{0}})dr.
\end{align}
We then perform the radial integration by using the method of integration by, parts,
\begin{align}
-2\pi \int_{r=0}^{r=R} r^{2}(\frac{\partial}{\partial r}) (p+\frac{|\textbf{B}|^{2}}{2\mu_{0}})dr = \notag \\ -2\pi((p_{b} + \frac{|\textbf{B}|^{2}}{2\mu_0})R^{2} - \int_{r=0}^{r=R} (p+\frac{|\textbf{B}|^{2}}{2\mu_{0}})(2r)\, dr).
\end{align}

So the first integral on the right-hand side of Equation \eqref{int1} becomes, after integration,
\begin{equation}\label{surf}
-2\pi(p_b + \frac{|\textbf{B}|^{2}}{2\mu_0})R^{2} +4\pi \int_{r=0}^{r=R} (p+\frac{|\textbf{B}|^{2}}{2\mu_{0}})r\,dr
\end{equation}
where $p_{b}$ is the pressure at the boundary, which is zero providing we have no surface currents. This is justified with the smooth matching of the magnetic field at the boundary.

Together with the simplifying assumption $B_z=0$ at the boundary, we can rewrite Equation \eqref{surf} in the following form:
\begin{align}\label{use2}
-2\pi(\frac{|\textbf{B}_{\phi}|^{2}}{2\mu_0})R^{2} +4\pi \int_{r=0}^{r=R} (p+\frac{|\textbf{B}|^{2}}{2\mu_{0}})rdr  = \notag \\-2\pi(\frac{|\textbf{B}_{\phi}|^{2}}{2\mu_0})R^{2} + 2((\gamma-1)U+\mathcal{M}_{\phi} +\mathcal{M}_{z}),
\end{align}
where $U$ is the thermal energy of motion per unit length, $\gamma$ is the ratio of specific heats, and $\mathcal{M}$ is the total magnetic energy per unit length. The subscripts on $\mathcal{M}$ indicate the contributions from the magnetic field components, i.e., azimuthal and axial.

To carry out the last integral on the right-hand side of Equation \eqref{int1},
\begin{align}\label{use3}
-\int_{\phi=0}^{\phi=2\pi}  \int_{r=0}^{r=R} \frac{B_\phi^{2}}{\mu_{0}}rd\phi dr = -2\pi \int_{r=0}^{r=R} \frac{B_\phi^{2}}{\mu_{0}}rdr = \notag \\ -2\pi \int_{r=0}^{r=R} \frac{2B_\phi^{2}}{2\mu_{0}}rdr = -2\mathcal{M}_{\phi},
\end{align}
we substitute Equations \eqref{use1}, \eqref{use2} and \eqref{use3} into Equation \eqref{int1} and consider equilibrium with the acceleration term equal to zero. This leads, after some algebra, to the equation
\begin{equation}\label{almost1}
\Omega-2\pi\frac{{B}_{\phi}^{2}}{2\mu_0}R^{2}+2(\gamma-1)U + 2\mathcal{M}_{z} =0
\end{equation}
where  $B_\phi$ is evaluated on the surface of the cylinder at $r=R$, and
\begin{equation}\label{ge}
\Omega = -G\lambda^{2}
\end{equation}
denotes the gravitational energy per unit length of the cylindrical cloud. Interestingly, the azimuthal magnetic energy $\mathcal{M}_{\phi}$ due to the magnetic tension in Equation \eqref{use3} is canceled by a similar term with a minus sign due to the magnetic pressure in Equation \eqref{use2}.

By using Ampere's law
\begin{equation}\label{max2}
{B}_{\phi}^2= \left(\frac{\mu_{0} I}{2 \pi R}\right)^2
\end{equation}
in terms of the total current $I$ carried by the cylinder, we see that Equation \eqref{almost1} becomes
\begin{equation}\label{virial}
\Omega+2(\gamma-1)U + 2\mathcal{M}_{z} -\frac{\mu_{0}}{4\pi}\, I^{2} =0.
\end{equation}
Equation \eqref{virial} is our new virial equation, representing hydrostatic equilibrium in the presence of both axial and azimuthal magnetic fields.

It is worth noting that if we set the current to zero and remove the azimuthal magnetic field then we recover the virial equation
\begin{equation}
\Omega+2(\gamma-1)U+2\mathcal{M}_{z}=0
\end{equation}
due originally to Chandrasekhar \& Fermi (1953) for the case with only axial magnetic field.

Remarkably, the last term $-\frac{\mu_{0}}{4\pi}\, I^{2}$ of Equation \eqref{virial} featured in this new virial equation, arises from the boundary term when integrating the magnetic pressure. Indeed, the existence of this new contribution is dictated by Ampere's law for a current carrying filamentary cloud, and is absent from a finite three-dimensional cloud. Furthermore, its minus sign has an important implication that while the magnetic field tends to stabilize a finite three-dimensional cloud, see, e.g., Shu (1992), the azimuthal magnetic field can destabilize a filamentary cloud. This is to be expected  through the pinch effect of MHD, which we shall explore in the next section.

\section{GRAVITATIONAL STABILITY WITH AZIMUTHAL MAGNETIC FIELD ONLY}

Here we shall set the axial magnetic field to be zero and focus on the effect of the azimuthal magnetic field. Thus,
\begin{equation}\label{Mz0}
\mathcal{M}_{z} =0.
\end{equation}
The aim is to provide a simple estimate of the gravitational stability of the system based on the strength of the azimuthal magnetic field and mass density of the cloud. Therefore, we consider an idealized case where the mass density $\rho$ and current density $J$ are homogeneous throughout the medium.

The total energy per unit length of the system takes the form
\begin{equation}\label{E}
E = \Omega + U + \mathcal{M}_{\phi}
\end{equation}
where $\mathcal{M}_{\phi}$ is the azimuthal magnetic energy per unit length.
In our present estimate, we shall evaluate $\mathcal{M}_{\phi}$ using the azimuthal magnetic field interior to the cylinder:
\begin{equation}\label{Mphi}
\mathcal{M}_{\phi} = \int_{\phi=0}^{\phi=2\pi} \! \int_{r=0}^{r=R} \frac{B^{2}}{2\mu_{0}}\,rd\phi\,dr.
\end{equation}
by assuming that the azimuthal magnetic field is effectively canceled by that of the return current exterior to the cylinder (Falgarone \& Passot 2003).

The integral in Equation \eqref{Mphi} can be readily evaluated using the constant current density $J$ and the corresponding total current $I = \pi R^2 J$, yielding
\begin{equation}\label{J}
\mathcal{M}_{\phi}=\frac{\pi\mu_{0}}{16} R^4 J^2=\frac{\mu_{0}}{16\pi} I^2.
\end{equation}

Therefore, by applying Equations \eqref{Mz0} and \eqref{J}, the virial equation \eqref{virial} now becomes
\begin{equation}\label{k}
\Omega - 4\mathcal{M}_{\phi} + 2(\gamma-1)U=0.
\end{equation}

Eliminating $U$ from Equation \eqref{k} by using the energy expression \eqref{E} we see that
\begin{align}\label{k1}
2(\gamma -1)E=2(\gamma +1)\mathcal{M}_{\phi} + (2\gamma -3)\Omega.
\end{align}
As a necessary condition for the dynamical stability of the cylindrical cloud, its total energy per unit length must be less than zero ($E < 0$). From Equation \eqref{k1}, this means the following inequality:
\begin{align}\label{GM}
-(2\gamma -3)\Omega >2(\gamma +1)\mathcal{M}_{\phi}.
\end{align}

Therefore, from Equation \eqref{GM} we can reach a necessary condition for stable equilibrium, using, e.g., $\gamma=\frac{5}{3}$ for monatomic plasma, as follows:
\begin{align}\label{GM1}
-\frac{1}{4}\Omega>\mathcal{M}_{\phi}.
\end{align}
This places an upper limit on the azimuthal magnetic energy to be a quarter of the absolute gravitational energy for the system to be gravitationally stable.

By using Equations \eqref{ge}, \eqref{J} and the expression for mass per unit length $\lambda=\pi R^{2} \rho$ the inequality \eqref{GM1} becomes
\begin{align}
\frac{1}{4}G(\pi R^{2} \rho)^{2}>\frac{\mu_{0} \pi J^{2}R^{4}}{16},
\end{align}
leading to the conclusion that for gravitational stability the current density must necessarily satisfy the following inequality:

\begin{equation}\label{stability}
J< \sqrt[]{\frac{4\pi G}{\mu_{0}}} \rho.
\end{equation}

\section{DISCUSSIONS}

The new form of virial equation \eqref{virial} gives us a gravitationally bound condition for the filament in hydrostatic equilibrium, this alone however may not mean that the filament is stable. The result can be reduced to the one derived by Chandrasekhar \& Fermi (1953) by setting the current (and hence the azimuthal magnetic field) to zero. This is an important result because it demonstrates that the azimuthal magnetic field significantly modifies the previously established necessary condition for gravitational stability. However, in making this comparison we need to be cautious in comparing our result with their findings. The azimuthal magnetic field can extend to infinity whereas the axial magnetic field is restricted to existing solely inside the plasma and falling to zero at the boundary but it need not be subject to such a restriction. It is possible for the axial magnetic field to exist outside the plasma and to be non-zero at the boundary. The assumption which is made that the pressure of the plasma is zero at the boundary need not be the case if surface currents exist. This would create a Lorentz force at the surface inducing pressure at the boundary. These additional considerations will no doubt affect the gravitational stability of the system and should be accounted for. On the other hand even without these considerations we can now say without doubt that an azimuthal magnetic field does indeed influence the gravitational stability of the system, proportionally to the current squared.

Removing the axial magnetic field, leaving the azimuthal magnetic field, we derive the gravitational stability condition given by Equation \eqref{stability}. This inequality \eqref{stability} sets up an estimated upper limit of the current density beyond which gravitational instability occurs, for given density of the cylindrical filament. For example, using a relatively low density of 10$^4$cm$^{-3}$ at 10 K (Henning et al 2010) and the radius of 0.15 lt-yr (Arzoumanian et al. 2011), whose values are typical of such filaments, we see that under these conditions, inequality \eqref{stability} yields:
\begin{equation}
J< 4.4\times 10^{-19}\text{Am}^{-2}.
\end{equation}
In this case, it is clear that the current density on the right-hand side of the above inequality physically represents a few electrons per second passing through a meter squared of area. Using the given radius, this value is equivalent to the minimum current of $2.8\times 10^{12}$A to violate gravitational stability.

The above current generates an azimuthal magnetic field of $3.9 \mu$G, at the boundary of the cylinder, which is around the expected strength of magnetic field. For example, the local ISM has a magnetic field strength of $4\mu$G (Opher et al. 2009). The Radio Arc filaments have been estimated to have a magnetic field strength of $10\mu$G (Ferri\`{e}re 2009). Let us examine the Serpens South Cluster, which is 848 lt-yr from Earth. It is in a very early part of the star formation process. It is emitting near-infrared light detected by \textit{Herschel} (Bontemps et al. 2010) and NASA's \textit{Spitzer Space Telescope} (Gutermuth et al. 2008).

This IRDC is predicted to have a magnetic field of a few hundred micro Gauss in two different zones along the main filament (Sugitani et al 2011). According to our analysis, it would be gravitationally unstable.  Consequently, small perturbations around its MHD equilibrium may be subject to dynamical instability, thereby growing in an unbound manner, while the stable parts of the filament remain as fragments. This fragmentation could be followed by the eventual collapse of the resulting fragments, trigging star formation, in fashion of Jeans' instability. Such processes have been demonstrated through MHD numerical simulations (Fiege $\&$ Pudritz 2000a,2000b) and semi-analytical solutions for self-similar reduced models of MHD (Hennebelle 2003), where certain perturbative oscillations of wavelength greater than some scale related to the azimuthal magnetic field have been observed. This scenario could be relevant for the observation of the above IRDC, where the infra-red detectors show, at its core, a bright cluster with more stars forming in a dispersed manner along the body of the cylinder. There are 35 protostars in total beginning to form (Bontemps et al. 2010).

The idealistic magnetic fields which we have considered can be expanded to consider a more complex magnetic field geometry, perhaps as a result of turbulence (Federrath et al. 2011), it can be shown that weak magnetic fields can be exponentially amplified through small dynamo effects during the process of star formation (Sur et al 2010). Such structures producing tangled magnetic fields can affect stability, e.g. by making the molecular cloud more difficult to compress.

We have adopted throughout the cylinder to be infinitely long to provide a first approximation to filamentary clouds with large aspect (length to width) ratio. However, it is also possible to extend this model to a closed filamentary structure. This is where we have a finite radius but a feedback mechanism for the current within the filament. In this case, there would exist the current inside the molecular cloud producing the azimuthal magnetic field but in addition there would exist a return current outside of the cloud. The form of this return current would depend on whether the molecular cloud was isolated or existed in a much larger network, both of these possibilities have been briefly explored (Falgarone \& Passot 2003). If a finite cylinder is considered higher modes will be excited, requiring a more in-depth analysis.

\acknowledgments

The authors are most grateful to M. Griffin (Cardiff and Herschel-SPIRE) and G. White (Open University and RAL) for helpful discussions. J.McL. acknowledges the Cormack Committee of the Royal Society of Edinburgh for financial support. C.W. and R.B. thank the STFC Centre for Fundamental Physics for partial support.

\end{document}